\documentclass[12pt,a4paper,twoside]{article}

\usepackage{graphicx,amssymb,amsmath,amsfonts}
\usepackage{epsfig}
\usepackage{xcolor}
\usepackage[mathscr]{eucal}
\usepackage{comment}
\usepackage{bm}
\usepackage[version=4]{mhchem}

\usepackage[sort&compress,numbers]{natbib}
\usepackage{doi}
\usepackage{hyperref}

\hypersetup{
    bookmarks=true,         
    unicode=false,          
    pdftoolbar=true,        
    pdfmenubar=true,        
    pdffitwindow=false,     
    pdftitle={Certificate},    
    pdfauthor={valiska},     
    pdfsubject={},   
    pdfcreator={valiska},   
    pdfproducer={},  
    pdfkeywords={}, 
    pdfnewwindow=true,      
    colorlinks=false,       
    linkcolor=red,          
    citecolor=green,        
    filecolor=magenta,      
    urlcolor=cyan           
}
\def\TheAuthor{Svyatoslav Kondrat}                                   

\def\TheTitle{Physics and modelling of intracellular diffusion}                        

\def\TheHeading{Physics and modelling of intracellular diffusion}                                
\def\TheInstitute{Department of Complex Systems\newline
Institute of Physical Chemistry
}                               
\def\TheAddress{Warsaw, Poland}            
\def\ThePart{D}                                             
\def\TheChapter{1}                                          
\include{defines}

\usepackage{xspace}
\newcommand{\latin}[1]{{\it #1}}
\newcommand{\ie}{\latin{i.e.}\@\xspace}
\newcommand{\eg}{\latin{e.g.}\@\xspace}

\newcommand{\etc}{\latin{etc.}\@\xspace}
\newcommand{\viz}{\latin{viz.}\@\xspace}

\def\la{\langle}
\def\ra{\rangle}

\newcommand{\fig}[1]{Fig.~\ref{#1}}

\newcommand{\Fig}[1]{Figure~\ref{#1}}

\newcommand{\figtitle}[1]{\textbf{#1}}

\newcommand{\sect}[1]{Section~\ref{#1}}
\newcommand{\sects}[1]{Sections~\ref{#1}}
\newcommand{\chapt}[1]{{Chap.~#1} of this book}

\newcommand{\eq}[1]{eqn~(\ref{#1})}
\newcommand{\eqs}[1]{eqn~(\ref{#1})}
\newcommand{\Eq}[1]{Equation~(\ref{#1})}
\newcommand{\Eqs}[1]{Equations~(\ref{#1})}

\newcommand{\myref}[1]{Ref.~\cite{#1}}

\newcommand{\rr}{\boldsymbol{r}}

\newcommand{\vv}{\boldsymbol{v}}
\newcommand{\x}{\boldsymbol{x}}
\newcommand{\I}{I}

\newcommand{\MSD}{\mathrm{MSD}}
\newcommand{\Rand}{\boldsymbol{\xi}}
\newcommand{\RandAv}{\boldsymbol{R}}

\newcommand{\Force}{\boldsymbol{F}}

\usepackage{ferienschule_18}                                  

\makeindex

\begin{document}

\MakeTitel           
\tableofcontents     

\vfill
\rule{42mm}{0.5pt}\\
{\footnotesize appeared in: \textit{Physics of Life}, $49^{{\rm th}}$ IFF Spring School 2018, Series: Key Technologies, Vol.~158, Forschungszentrum J{\"{u}}lich Publishing.}

\newpage

\section{Introduction}

An important difference between a typical (non-biological) laboratory system and a living cell is an enormous variety and amount of metabolites and macromolecules diffusing, interacting and reacting in cells. Laboratory-scale macromolecular experiments often deal with volume fractions of the order of a few percents, while the volume taken up by the macromolecules in a biological cell is of the order of $20\%$ to $50 \%$~\cite{zimmerman:jmb:91:ecolicrowd} (Figure~\ref{fig:crowded}). The effects of such a crowded environment on macromolecular structure and reactivity inside living cells were first analysed in the early 80s~\cite{minton:biopol:81, fulton:cell:crowd:82}, but its importance for diffusion and reactions had not been appreciated until recently. For instance,  \citeauthor{ellis:cosb:crowd:01} wrote, concluding his 2001 review article \cite{ellis:cosb:crowd:01}, that crowding \textit{``should become a routine variable to study''} and continued suggesting journals to \textit{``reject manuscripts on the grounds that this important variable has not been controlled.''}

Most physicochemical processes proceed differently in a biologically crowded environment\iffindex{crowding}: Diffusion slows down enormously and reactions may occur with different rates, in particular due to the reduced diffusion. Understanding diffusion in dense biological systems is, therefore, of critical importance for life sciences and for designing biotechnological applications. In this Lecture, I will introduce the notion of Brownian motion (or self-diffusion) and we will discuss briefly how to describe, model and measure the diffusion properties (we shall restrict our attention to translational diffusion, however). The focus will be on physics and simulations, with a particular emphasis on the effects important for crowded, biologically relevant systems. Various aspects of crowding are also discussed in \chapt{D2}.

\begin{figure}[!t]
    \begin{center}
    	\includegraphics*[width=0.3\textwidth]{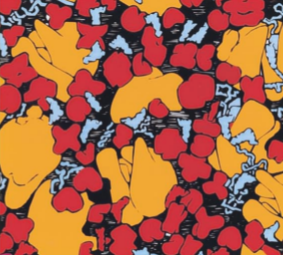}
	\includegraphics*[width=0.273\textwidth]{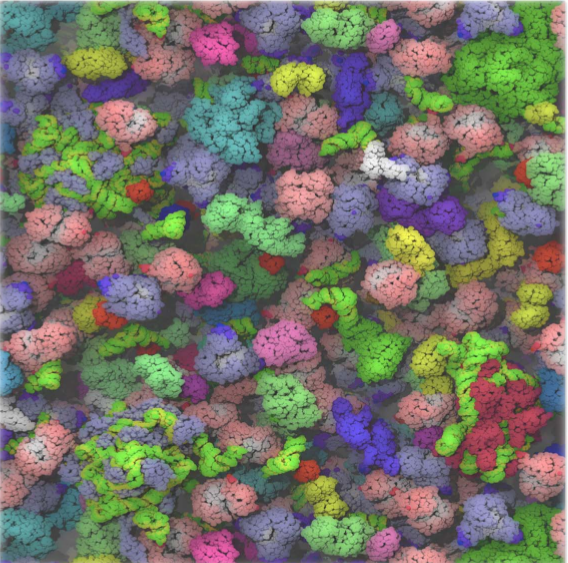}
        \caption{
        \label{fig:crowded}
	    \figtitle{Crowded environment inside living cells.} (left) An artist's view of the cytoplasm. Picture taken from~\cite{ellis:cosb:crowd:01}. (right) Snapshot from the Brownian dynamics simulations of Ref.~\cite{mcguffee:plos:10}. Various macromolecules are depicted by different colors.
	}
\end{center}
\end{figure}

In this short Chapter it has not been possible to account for all relevant approaches and phenomena important for intracellular diffusion. I thus refer the interested readers to a few reviews \cite{elcock:curopstructbio:10, dlugosz:bmcbiophys:11, sholnick:jcp:16:Perspective, feig_sugita:jpcb:17}, and note that this contribution shall merely be considered as a humble introductory note into the physics and modelling of diffusion in biophysically crowded environments inside living cells.

\section{Macromolecular self-diffusion}
\label{sec:diff}

\iffindex{self-diffusion} 
\iffindex{Brownian motion|see {self-diffusion}} 

Brownian motion or self-diffusion is the random motion of macromolecules (or other objects) suspended in a fluid. Such motion results from the collision of macromolecules with the fast-moving smaller molecules in a suspension, such as water and metabolites. This process shall be distinguished from a related process of diffusion, sometimes called transport diffusion, which is the natural motion of molecules from a region of high concentration to a region of low concentration (more precisely, high and low chemical potentials). Here, we shall deal exclusively with self-diffusion of macromolecules, particularly in a crowded environment inside living cells, but we shall briefly mention the transport diffusion in conclusions (\sect{sec:concl}). For brevity, however, we shall use self-diffusion and diffusion interchangeably where it does not lead to confusion.

\subsection{How to characterize diffusion?}

One way to quantify self-diffusion is to look at the mean square displacement (MSD) of a diffusing particle (the mean displacement is obviously zero): \iffindex{mean square displacement (MSD)} 
\begin{align}
\MSD(t) = \langle \left[\rr(t) - \rr(0)\right]^2 \rangle,
\end{align}
where $\langle \cdots \rangle$ means \emph{ensemble} averaging, $\rr(t)$ is the molecule's position at time $t$ and $t=0$ is the beginning of observation. In practice the number of trajectories is often very limited, and \emph{time} averaging  within a small time frame is added to improve statistics (provided the system behaves ergodically, which is not always the case~\cite{ghosh_metzler:pccp:15, ghosh_metzler:njp:16}).

\begin{figure}[!t]
    \begin{center}
    	\includegraphics*[width=0.6\textwidth]{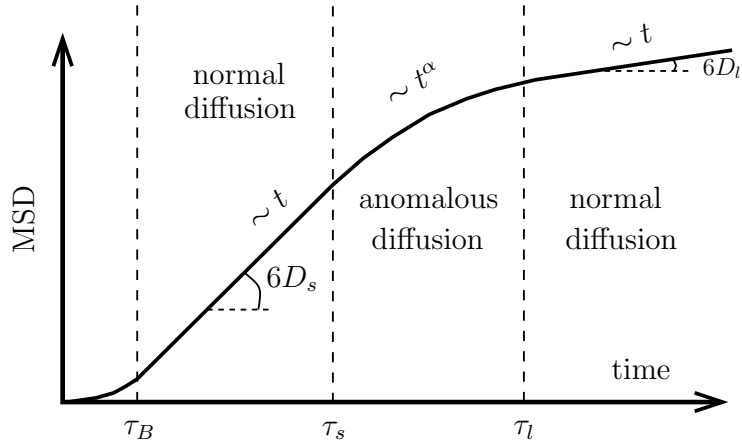}
        \caption{
        \label{fig:msd}
	\figtitle{Normal and anomalous diffusion.} Schematic of the behaviour of the mean-square displacement (MSD) with time. $D_s$ and $D_l$ are short-time and long-time self-diffusion coefficients, respectively, $\tau_B$ is the timescale of momentum relaxation, and $\tau_s$ and $\tau_l$ are the crossover times from the short-time and to the long-time normal regimes. 
	}
\end{center}
\end{figure}

At very short times, $\MSD \sim t^2$ due to  momentum relaxation ($t< \tau_B \sim m/\gamma$, where $m$ is the particle mass and $\gamma$ the friction coefficient; see \fig{fig:msd}). This time scale is commonly omitted when discussing Brownian motion.

For a dilute system, the MSD typically behaves linearly with time for $t > \tau_B$; then \iffindex{diffusion coefficient}
\begin{align}
	\label{eq:Ds}
	D =\lim_{t\to \infty}  \frac{\MSD(t)}{2dt}  
\end{align}
defines the diffusion coefficient (here $d$ is the dimensionality). The diffusion coefficient is one of the most important quantitative measures of molecular diffusion. It is analogous to the velocity in the classical mechanics and tells us the (approximate) distance a particle can travel in time $t$, which is $l \approx \sqrt{D t}$.  The diffusion coefficient can be extracted from simulations (\sect{sec:sim}) and assessed experimentally (\sect{sec:measure}).

In a crowded system (such as the cytoplasm) the situation is more complex. In this case the MSD does not always behave linearly with time, and one can distinguish short-time and long-time self-diffusion coefficients, $D_s$ and $D_l$, respectively, as schematically depicted in \fig{fig:msd}. In between these two \emph{normal} regimes\iffindex{self-diffusion!normal}, \ie, for $\tau_s < t < \tau_l$, there is a region of \emph{anomalous} diffusion\iffindex{anomalous subdiffusion|see {self-diffusion, anomalous}}\iffindex{self-diffusion!anomalous}. The crossover time to the anomalous diffusion can be roughly estimated from the mean inter-macromolecular distances, assuming the anomalous regime commences on the time scale associate with the macromolecular collisions~\cite{kondrat:pb:15:bdcomp}. For macromolecules of the same radius $a$ such simple considerations give\iffindex{self-diffusion!crossover to anomalous regime}
\begin{align}
	\label{eq:tau_s}
	\tau_s \approx \frac{a^2}{D_s} \left([4\pi /(3\eta)]^{1/3} - 2\right)^2,
\end{align}
where $\eta$ is the packing fraction of macromolecules. \Eq{eq:tau_s} shows that $\tau_s$ increases with the size of a macromolecule ($D_s$ decreases with $a$), and it decreases monotonically with increasing the volume fraction. Specifically, $\tau_s \to \infty$ as $\eta \to 0$, as one may expect, and $\tau_s$ vanishes at $\eta_\mathrm{max} =\pi/6 \approx 0.54$, manifesting that at high volume fractions a macromolecule encounters other macromolecules on the length scale of its own dimension; note that $D_s$ depends on $\eta$ due to hydrodynamic interactions (which, in general, contain many-body far and near-field contributions), but here we neglect this dependence for simplicity. This simple estimate turns out to be in a good agreement with simulations, at least in some range of volume fractions~\cite{kondrat:pb:15:bdcomp}. Taking as an example $a = 2.5$nm and the diffusion coefficient $D_s=6.6$\AA$^2$/ns (which corresponds to the tRNA/triphosphate isomerase), we find $\tau_s \approx 55$ns  for volume fraction $20\%$ (density $\approx 5$mM).

In the anomalous regime, one can define a time-dependent diffusion coefficient
\begin{align}
	\label{eq:anomal}
	D(t) = \frac{\MSD(t)}{2d t} = \Gamma t^{\alpha-1},
\end{align}
where $\alpha$ is the exponent of anomaly \iffindex{anomaly exponent}and $\Gamma$ is the generalized transport (or anomalous diffusion) coefficient\iffindex{generalized transport coefficient|see {anomaly exponent}}. For a crowded system, such as the cytoplasm, $\alpha < 1$; this is termed anomalous \emph{subdiffusion}; in the opposite case, when $\alpha > 1$, it is called anomalous superdiffusion.

The crossover time to the long-time normal diffusion, $\tau_l$, is not easy to estimate, but experiments and simulations suggest that $\tau_l$ is in the range from tens of microseconds to milliseconds~\cite{vilaseca_mas:tca:11, kondrat:pb:15:bdcomp, blanco_mas:entropy:17}. It is also possible that $\tau_l \to \infty$~\cite{hoefling:rpp:anomal:13}, or that the system size is too small (\viz, the linear size $L < \sqrt{D_l \tau_l}$), in which case the long-time normal regime is not observed. 

In addition to MSD and two diffusion coefficients, there are other quantities characterizing diffusion, such as rotational diffusion coefficients, mean dwell times, time autocorrelation functions, \etc However, for the present discussion, it will be sufficient to consider MSD and $D_{s,l}$.

\section{How to model diffusion?}
\label{sec:sim}

Ideally, one would like to perform `\textit{ab initio}' simulations of a whole system, in order to approximate the reality as close as possible. Within the classical approach, this amounts to solving Newtonian equations of motions for all molecules in a system (for details see \chapt{A10}). For realistic biological systems, however, such simulations are dramatically expansive computationally. Although molecular dynamics simulations of crowded cytoplasm-like environments do exist~\cite{feig_sugita:jpcb:17}, the time scales covered so far are of the order of a few nanoseconds to at most microsecond, which is often below both $\tau_s$ and $\tau_l$. 

Another way to deal with such systems is to separate short and long time scales, and to consider `averaged' equations of motions on longer time scales. This leads to the well-known Langevin equation, which we discuss below.

\subsection{Langevin equation}

\iffindex{Langevin equation}

In a biophysical context, a natural method of dividing a system into short and long time scales is to consider the motion of \emph{macromolecules} and to average out over the motion of small molecules, such as water, ions and (some) metabolites. Then the Langevin equation for $N$ spherically-symmetrical macromolecules is ($i=1,\cdots,N$)
\begin{align}
	\label{eq:Langevin}
	m_i \frac{d^2 \rr_i }{dt^2} = - \nabla_i W(\{\rr_k\}) - \sum_{j=1}^N \gamma_{ij} \frac{d \rr_j}{dt} + \sum_{j=1}^N \sigma_{ij} \Rand_j(t),
\end{align}
where $m_i$ is the mass of molecule $i$, $W$ is the \emph{effective} interaction potential, which takes into account solvent-mediated interactions, and $\nabla_i = \partial / \partial \rr_i$ is the Nabla operator, so that $\Force_i = - \nabla_i W$ is the effective force acting on molecule $i$. 

Now, $\gamma_{ij}$ is the $N\times N$ friction matrix (of $d\times d$ blocks, where $d$ is the dimensionality) and $\Rand_i$ is the \emph{stochastic} or fluctuating force, both due to the presence of solvent. The stochastic force is the white noise with zero mean, $\la \Rand_i \ra = 0$, which is uncorrelated,
\begin{align}
	\label{eq:RandomAutocorr}
	\la \Rand_i(t) \Rand_j (t') \ra = 2 \delta_{ij} \delta (t-t').
\end{align}
The coupling constants $\sigma_{ij}$, \ie, the magnitudes of the random forces, are related to the friction matrix by the fluctuation-dissipation theorem
\begin{align}
	\label{eq:FluctDiss}
	\gamma_{ij} = (k_B T)^{-1} \sum_k \sigma_{ik} \sigma_{kj},
\end{align}
where $k_B$ is the Boltzmann constant and $T$ kinetic temperature, as usual. Relation (\ref{eq:FluctDiss}) can be obtained by integrating \eq{eq:Langevin} once and calculating the average of the velocity squared, $\la v^2\ra$, and then employing the equipartition theorem $ m_i\la v_i^2\ra/2 = d k_BT/2$.

The Newtonian equations of motions (no solvent) are reproduced by setting $\sigma_{ij} = 0$ (and hence $\gamma_{ij} = 0$) and, since $W$ is the effective potential, by replacing $W$ by the interaction potential in the absence of solvent.

At low Reynolds numbers, \ie, for a slowly moving, viscous fluid, momentum relaxations can be omitted at long time scales ($t \gg \tau_B$, \fig{fig:msd}). This means $d\vv_i/dt \ll \sum_j(\gamma_{ij}/m_i) \vv_j$, where $\vv_i=d\rr_i/dt$ is the velocity, and we obtain
\begin{align}
	\label{eq:BD}
	\frac{d \rr_i}{dt} = - (k_BT)^{-1} \sum_j D_{ij} \nabla_j W(\{\rr_k\}) + (k_BT)^{-1} \sum_{jk} D_{ik}\sigma_{kj} \Rand_j(t),
\end{align}
where $D_{ij} = k_BT(\gamma^{-1})_{ij}$ is the diffusion matrix, which is inverse of the friction tensor. \Eq{eq:BD} is the main equation for Brownian motion.

\subsection{Ermak-McCammon equation}
\label{sec:EM}

\iffindex{Ermak-McCammon equation}

Instead of applying the strong friction limit, as above, one can integrate the Langevin equation directly on the time scales larger than the momentum relaxation ($t \gg m_i/\gamma_{ii}$). Assuming that the diffusion matrix is position dependent, but varies slowly in space, it is possible to obtain for the particle displacement~\cite{ermak_mccammon:jcp:78}
\begin{align}
	\label{eq:EM}
	\Delta \rr = \rr(\Delta t) - \rr (0) 
		=  \Delta t\sum_j \nabla_j D_{ij}^{(0)} 
		-  \Delta t (k_BT)^{-1} \sum_j D_{ij}^{(0)} \nabla_j W^{(0)}
		+ \RandAv_i(\Delta t),
\end{align}
where the upper index in $D_{ij}^{(0)}$ and $W^{(0)}$ indicates that the diffusion matrix and the potential must be evaluated at $t=0$ (in case they depend on time and/or positions of the macromolecules); $\RandAv_i(\Delta t)$ is a random displacement averaged over time $\Delta t$, which satisfies~\cite{ermak_mccammon:jcp:78}
\begin{align}
	\label{eq:EM:Covar}
	\la \RandAv_i(\Delta t)  \RandAv_j(\Delta t) \ra = 2 D_{ij}^{(0)} \Delta t.
\end{align}
\Eqs{eq:EM} and (\ref{eq:EM:Covar}) have been derived by \citeauthor{ermak_mccammon:jcp:78} \cite{ermak_mccammon:jcp:78} and can be viewed as a finite-difference (propagator) scheme for Brownian dynamics simulations. This scheme is equivalent to the first-order Euler algorithm for ordinary differential equations, and has been extended by \citeauthor{iniesta:jcp:90}~\cite{iniesta:jcp:90} to the second order Runge-Kutta approach by taking into account the second-order corrector step.

The \citeauthor{ermak_mccammon:jcp:78}~\cite{ermak_mccammon:jcp:78} derivation sets naturally the limits on the time step $\Delta t$. Indeed, on the one hand, it is clear that $\Delta t$ must  be greater than the timescale of the momentum relaxation, \ie, $\Delta t \gg m_i/\gamma_{ii}$. On the other hand, $\Delta t$ must be sufficiently small such that the interaction potential and the friction matrix can be considered constant during the time step $\Delta t$. 

For a single spherical particle in a homogeneous media \eq{eq:EM} simplifies to
\begin{align}
	\label{eq:EM:1}
	\Delta \rr =  \RandAv_i(\Delta t) = \sqrt{2D_0 \Delta t}\; \x,
\end{align}
where $\x$ is a random vector satisfying the Gaussian distribution, so that $\la \RandAv^2(\Delta t)\ra = 2 D_0 \Delta t$. Using now \eq{eq:EM:1} to calculate particle's trajectory, and performing a sufficient number of independent simulations to gather enough statistics, we can calculate the MSD and extract the diffusion coefficient $D$ using \eq{eq:Ds} to confirm that it coincides with the diffusion coefficient $D_0$ used as an input in \eq{eq:EM:1}.

For $N$ spherical macromolecules in a homogeneous medium one has
\begin{align}
	\label{eq:EM:N}
	\Delta \rr =  
		-  \Delta t (k_BT)^{-1} \sum_j D_{ij} \nabla_j W^{(0)}
		+ \RandAv_i(\Delta t).
\end{align}
The random displacement is
\begin{align}
	\label{eq:RandomForce}
	\RandAv_i = \sqrt{2 \Delta t} \sum_j B_{ij} \cdot \x_j,
\end{align}
where $\x_j$ is a Gaussianly distributed random vector, the dot means a convolution over $d$ components of $\x_j$ (note that $B$ is a $N\times N$ matrix of  $d\times d$ blocks), and
\begin{align}
	\label{eq:D_BB}
	D_{ij} = \sum_k B_{ik}B_{kj}.
\end{align}
Thus, in order to calculate the random force one needs to take a `square root' of the diffusion matrix. Since essentially any $B$ that satisfies \eq{eq:D_BB} can be used in \eq{eq:EM:N}, one often takes the standard Cholesky decomposition~\cite{goloub_loan:matrix:book:96}. In the case discussed, however, the matrix $D$ is diagonal hence the `decomposition' can be performed in a straightforward way and only once before a simulation starts (assuming that the diffusion constants do not change in time). However, this changes when the hydrodynamic interactions are taken into account. We address this in the next section.

\subsection{Hydrodynamic interactions}

\iffindex{hydrodynamic interactions}

As we have discussed, in Brownian dynamics simulations the water, metabolites and other small molecules inside a cell are not taken into account explicitly, but effectively present a viscous environment for diffusion of macromolecules. A  macromolecule moving in this viscous medium excites a long-range flow that affects other molecules, resulting in an effective interaction between the macromolecules. 

In order to account for such \emph{hydrodynamic interactions}, one needs to solve the Navier-Stokes equation for the solvent for current positions and velocities of macromolecules each simulation step. This can be done by calculating the viscous stress $\gamma_h$ exerted on a macromolecule by other macromolecules and noticing that the force acting on this macromolecule is $\gamma_h \vv$, where $\vv$ is its velocity. This permits us to incorporate the hydrodynamic interactions into the diffusion matrix $D_h = (k_BT) \gamma_h^{-1}$. Assuming that such a viscous stress is a superposition of independent contributions from each macromolecule, and further neglecting the effects related to the macromolecule sizes, which shall be valid for large distances, it is possible to obtain for macromolecules of the same radius $a$~\cite{rotne:jcp:69, yamakawa:jcp:70}
\begin{align}
	\label{eq:RPY}
	D_{ij} = \frac{k_B T}{8\pi \nu r_{ij}}\left\{
		\I + \frac{\rr_{ij} \otimes \rr_{ij}}{r_{ij}^2}
			+ \frac{2a^2}{r_{ij}^2}\left( \frac{1}{3} \I - \frac{\rr_{ij} \otimes \rr_{ij}}{r_{ij}^2}\right)
		\right\},
\end{align}
where $\rr_{ij} = \rr_i - \rr_j$ is a vector connecting the centers of macromolecules $i$ and $j$, $\otimes$ denotes tensor product, $\I$ is the $3 \times 3$ unit matrix, and $\nu$ is the fluid viscosity; the diagonal components are $D_{ii} = (k_B T/6\pi \nu a)\I$. \Eq{eq:RPY} is known as the Rotne-Prager-Yamakawa (RPY) tensor\iffindex{Rotne-Prager-Yamakawa (RPY) tensor}. It can be extended to macromolecules of different sizes~\cite{delaTorre_Bloomfield:bioplo:77}, and, formally, to distances $r_{ij} < 2a$~\cite{rotne:jcp:69, yamakawa:jcp:70}. At short distances, however, also the lubrication and many-body forces must be taken into account~\cite{sholnick:jcp:16:Perspective}. To avoid this complication, a macromolecule can be approximated by small spheres placed on its surface~\cite{dlugosz:antosiewicz:jpcb:15, swan_wang:pf:16, dlugosz_antosiewicz:jpcb:16, antosiewicz_dlugosz:jcpb:17}; then the RPY tensor between these small spheres is valid in a wide range of macromolecular separations, but the tensor size increases proportionally to the number of spheres taken to approximate the macromolecules.

In any case, at \emph{each} Brownian dynamics (BD) simulation step the diffusion matrix must be factorized in order to obtain the hydrodynamically correlated random displacements (see \eqs{eq:RandomForce} and (\ref{eq:D_BB})). In addition, since such hydrodynamic forces are long-ranged (\eq{eq:RPY}), it is necessary to take into account not only the macromolecules present in a computational box, but also all their periodic images (in simulations of a bulk system). This can be done by using the so-called Ewald summation~\cite{beenakker:jcp:86:EwaldRP}. All this increases the computational cost of BD simulations significantly. For instance, the frequently used Cholesky decomposition, which we have already mentioned, scales as $N^3$, where $N$ is the number of macromolecules; the pair-wise inter-macromolecular interactions scale as $N^2$, and the Ewald summation as $N^2$ or $N\log(N)$, depending on the method used; thus the hydrodynamics are the bottleneck of BD simulations. Their computational cost can be lowered by using Chebyshev approximation for the diffusion matrix developed by~\citeauthor{fixman:macromol:86}~\cite{fixman:macromol:86}, but it still scales as $N^{2.5}$~\cite{dlugosz:bmcbiophys:11}. \citeauthor{geyer_winter:jcp:09}~\cite{geyer_winter:jcp:09} have proposed an approximate $N^2$ algorithm, which is based on a certain Ansatz (expansion) for the random force, with the unknown parameters determined approximately from the appropriate variance-covariance relation akin of \eq{eq:RandomAutocorr}. It shows a good agreement with the standard approaches, at least for the tested systems~\cite{geyer:bmcbiophys:11:brownmove, schmidt_delaTorre:jcp:11}. There are also other important methods and improvements~\cite{benchio:brady:jcp:03, ando_chow_shkolnik:jcp:13, ando_shkolnik:jcp:12:Krylov, saadat_khomami:jcp:14, saadat_khomami:pre:15}. Of particular interest is the Stockesian dynamics~\cite{benchio:brady:jcp:03}, which allows to take into account both long-range many body and short-range lubrication forces. Discussion of these methods is beyond the scope of this Chapter, but various approaches to hydrodynamics are discussed in \chapt{A10}.

\subsection{Software packages}

\iffindex{software packages!Brownian dynamics simulations}

There are a few open-source packages that offer ready-to-use software codes for Brownian dynamics (BD) simulations. The Brownian and Langevin dynamics have been implemented in the standard Gromacs~\cite{gromacs:url} and LAMMPS~\cite{lammps:url} simulation packages. However, Gromacs does not not seem to include the hydrodynamic forces, essential for crowded systems, as we shall see (\sect{sec:diff:simres}). In LAMMPS~\cite{lammps:url}, the hydrodynamics are implemented using lattice-Boltzmann approach. Brownmove~\cite{brownmove:url} is an implementation of the \citeauthor{geyer_winter:jcp:09} $N^2$ algorithm~\cite{geyer_winter:jcp:09}. BDpack~\cite{bdpack:url} is a software package that implements the recently introduced matrix-free method of \citeauthor{saadat_khomami:pre:15}~\cite{saadat_khomami:pre:15} for dilute and semi-dilute polymeric solutions. BD\_BOX~\cite{bdbox:url} is probably most versatile and highly optimized Brownian dynamics simulation package for rigid and flexible molecules~\cite{dlugosz:jcc:11}, but it takes into account only the long-range hydrodynamics. The GPU-optimized HOOMD-blue simulation toolkit~\cite{hoomd-blue:url} can also run BD simulations, and the hydrodynamics can be added via their plug-in system (\eg, with RPY tensor as in \myref{varga_swan:sm:15}). The short and long range hydrodynamics have recently been implemented in ESPResSo simulation package~\cite{ESPResSo:url}, but its performance has not been optimized (Christian Holm, personal communication).

Concluding, it seems that a well-developed and optimized open-source software package for BD simulations, which supports both long-range and short-range hydrodynamics, is not currently available.

\section{How to measure diffusion?}
\label{sec:measure}

Although the focus of this Chapter is physics and modelling of intracellular diffusion, in order to understand better the connection with experiments, we shall briefly mention the three main techniques developed for measuring diffusion properties. For details of these (fluorescent-based) methods see \chapt{A3}.

Perhaps conceptually the simplest (but technically advanced) is the \emph{single-particle tracking} (SPT)\iffindex{single-particle tracking (SPT)}. Such experiments amount to introducing a fluorescent dye into traced macromolecules and video-recording their diffusion. The MSD can be extracted from the recorded videos by analysing tracer trajectories, similarly as in simulations. Typically the spatial resolution of SPT is of the order of a few nanometres, with the time resolution on the scale of milliseconds, but high-speed tracking techniques exist allowing for the resolution of tens of microseconds~\cite{kusumi_etal:arbbs:05, greenleaf:arbbs:07:SPT}.

Probably the most frequently used method is the \emph{fluorescent correlation spectroscopy} (FCS)\iffindex{fluorescent correlation spectroscopy (FCS)}. Similarly as SPT, it relies on labeling tracers by a fluorescent dye, but in FCS the fluctuating fluorescent light intensity is measured, rather than the trajectories. The measured intensity can be related to the time autocorrelation function, from which the diffusion coefficient and the anomaly exponent can be extracted by fitting to the analytical expressions~\cite{krichevski_bonnet:rpp:02:FCS, hess_eta:biochem:02:FCS}. The advantage of FCS is the time resolution, which can be of the order of microseconds.

The \emph{fluorescent recovery after photobleaching} (FRAP) is similar to FCS\iffindex{fluorescent recovery after photobleaching (FRAP)}. Here, however, a small region in a sample with fluorescing macromolecules is initially bleached by an intense laser pulse, and the fluorescent light intensity (rather than fluctuations) is monitored as the bleached region recovers due to the diffusion of the fluorescent macromolecules from the outside of this region~\cite{reits:ncb:01:FRAP, lippincott_etal:natrev:01:ProtDynLivingCells,verkman:03:FRAP}.

All three methods are applicable \textit{in vivo} by in-cell expression of fluorescent macromolecules, frequently a green fluorescent protein (GFP)\iffindex{green fluorescent protein (GFP)}.

\section{Effect of crowding on long-time diffusion coefficients}

\subsection{Experiments}
\label{sec:diff:exp}

\begin{figure}
	\begin{center}
    	\includegraphics*[width=0.55\textwidth]{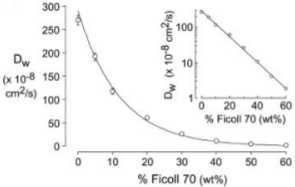}
	\end{center}
        \caption{
        \label{fig:diff:exp}
	\figtitle{Effect of crowding on diffusion from \textit{in vitro} experiments.} Diffusion coefficient of a rhodamine green as a function of the crowder concentration (Ficoll-70). The inset shows the same data in log scale. The figure shows the results of a FCS study. Reproduced from~\myref{dauty_verkman:jmr:04CrowdingSmallBig}.}
\end{figure}

\iffindex{intracellular diffusion!experiments}
Experiments indicate that the long-time diffusion coefficients of macromolecules are dramatically reduced in the crowded environments inside living cells. For instance, the diffusion coefficient of GFP \textit{in vivo} has been measured to be $10-15$ times lower than in a dilute solution~\cite{elowitz:jb:gfpmob:99, konopka:jb:gfpdiff:06}. 

A more systematic analysis can be carried out \textit{in vitro}, where the concentration of crowders can be easily controlled. \Fig{fig:diff:exp} shows the results of a FCS study of the long-time diffusion of rhodamine green in a concentrated solution of Ficoll-70 as crowders~\cite{dauty_verkman:jmr:04CrowdingSmallBig}. It has been found that the diffusion coefficient $D_w$ decreases exponentially with the crowder concentration $C$, \viz, $D_w = D_w^0 \exp \{- a C\}$, where $a$ is a fitting parameter and $D_w^0$ is the diffusion coefficient in infinite dilution. Interestingly, this work has also demonstrated that the reduction in diffusion is comparable for large macromolecules and for small solutes~\cite{dauty_verkman:jmr:04CrowdingSmallBig} (the plot not reproduced here).

\subsection{Simulations}
\label{sec:diff:simres}

\iffindex{intracellular diffusion!simulations}

\begin{figure}
	\begin{center}
	\includegraphics*[width=0.4\textwidth]{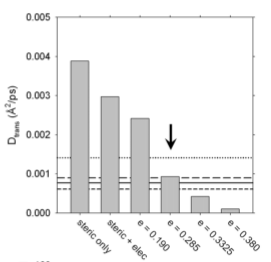}
    	\includegraphics*[width=0.55\textwidth]{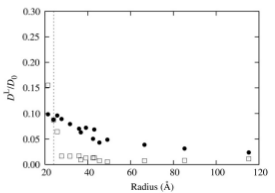}
	\end{center}
        \caption{
        \label{fig:diff:crowding}
	\figtitle{Effect of crowding on diffusion from Brownian dynamics simulations.} (left) Diffusion constant of GFP in the cytoplasm for different models (steric; steric and electrostatic; steric, electrostatic and Lennard-Jones with different interaction parameters). The arrow points out to the value of the depth of the Lennard-Jones interactions producing the observed 10-fold decrease of the diffusion coefficient. Reproduced from \myref{mcguffee:plos:10}. (right) Hydrodynamic interactions describe correctly the reduction of the diffusion coefficient in the cytoplasm (full circles), while the Lennard-Jones (van der Wals) interactions may overestimate it due to clustering (open rectangles). Thin vertical line denotes GFP. Reproduced from~\myref{ando:pnas:10}}
\end{figure}

From a modeling perspective, it shall be clear that at least short-range repulsive interactions between macromolecules must be take into account. \citeauthor{mcguffee:plos:10}~\cite{mcguffee:plos:10} have shown, however, that hard core (steric) interactions alone are not sufficient to reproduce the observed decrease of the diffusion coefficients. These authors considered a cytoplasm model consisting of $50$ different types of macromolecules (\fig{fig:crowded}), and studied the effect of steric, electrostatic and van der Waals interactions on diffusion. Using Brownian dynamic simulations, as described in \sect{sec:EM} (without hydrodynamic interactions), they showed that steric and electrostatic interactions are not sufficient to explain the observed slow-down of diffusion. However, the long-time self-diffusion coefficient turns out to be sensitive to the van der Waals interactions, and it seems possible to match the simulated $D_l$ for GFP with the diffusion coefficient measured in experiments by varying the depth of the Lennard-Jones potential within the physically reasonable range (\fig{fig:diff:crowding}). \iffindex{green fluorescent protein (GFP)}\iffindex{crowding}

However, \citeauthor{ando:pnas:10}~\cite{ando:pnas:10} have argued that van der Waals interactions may lead to clustering and overestimate the slow-down for larger macromolecules. In contrast, the hydrodynamic interactions (with short and long-range contributions) adequately reproduce the observed reduction of the diffusion coefficient (particularly of GFP) in the cytoplasm (\fig{fig:diff:crowding}).

Another important aspect of \citeauthor{ando:pnas:10}'s work~\cite{ando:pnas:10} is that the details of macromolecule's structure seem to be of negligible  importance for diffusion. This has been shown by comparing directly the translational diffusion coefficients in a molecularly-shaped system and in a system where the macromolecules were modeled as spherical particles.

\section{Anomalous subdiffusion}
\label{sec:diff:anomal}

\iffindex{intracellular diffusion!anomalous subdiffusion}

\begin{figure}
	\begin{center}
	\includegraphics*[width=0.4\textwidth]{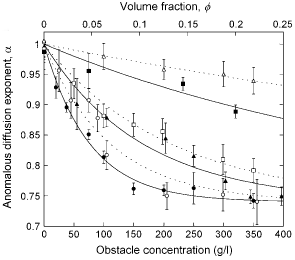}
    	\includegraphics*[width=0.15\textwidth]{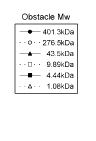}
    	\includegraphics*[width=0.4\textwidth]{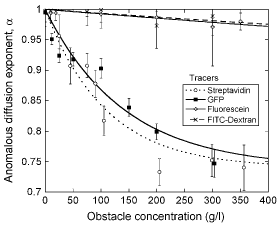}
	\end{center}
        \caption{
        \label{fig:diff:anomal}
	\figtitle{Anomalous subdiffusion from \textit{in vitro} experiments.} The exponent of anomalous subdiffusion, $\alpha$, as a function of the crowder (obstacle) concentration. (left) Exponent $\alpha$ for streptavidin in dextran of different sizes. (right) Exponent $\alpha$ for strepravidin and GFP in dextran 276.5 kDa, and for fluorescein and FITC-Dextran in 401.3 kDa dextran. The figure shows the results of a FCS study. Reproduced from~\myref{banks:bj:anomal:05}.}
\end{figure}

So far we have discussed diffusion focusing on its long-time normal behaviour. However, there is a `significant body of evidence' indicating that \textit{in vivo} diffusion is anomalous on extended time scales. \citeauthor{weiss:bj:crowdanomal:04} have even proposed to use the anomaly exponent\iffindex{anomaly exponent} ($\alpha$ in \eq{eq:anomal}) as a measure of how crowded a system is~\cite{weiss:bj:crowdanomal:04}. In their experiments, \citeauthor{weiss:bj:crowdanomal:04} studied the diffusion of dextran inside HeLa cells by FCS and found the anomaly exponent in the range between $\alpha = 0.73$ and $\alpha = 0.79$. Similar values have been obtained in other experiments~\cite{engelke_etal:pb:10, jeon_metzler:prl:11AnomalDiff}.

As the structural properties of the cell interior are not well known and easily controllable, a more systematic analysis can be carried out \textit{in vitro}, which allows to control the crowder's concentration and size. A FCS study shows~\cite{banks:bj:anomal:05}, in particular, that the anomaly exponent $\alpha$ decreases with increasing the crowder's volume fraction $\phi$ and saturates at $ \alpha \approx 0.75$ at high $\phi$ (\fig{fig:diff:anomal}); it can be well fitted by $\alpha (\phi) = \alpha_1 + \exp(-\phi/\phi_0)$, where $\phi_0$ and $\alpha_1$ are fitting parameters. The saturation is likely due to the entanglement of polymer chains of crowders (dextran in this case) at high concentrations so that effectively the tracer diffuses inside a cross-linked polymer network, which gives $\alpha = 3/4 \approx \alpha_1$ \cite{banks:bj:anomal:05}. Interestingly, the anomalous diffusion shows the same features for globular tracers (streptavidin and GFP), but the diffusion is normal or only slightly anomalous for polymer tracers (fluorescein and FITC-Dextran, \fig{fig:diff:anomal}), suggesting a different physics governing their diffusion~\cite{wang_pielak:jacs:10:ProtDiffInProt, shin_metzler:njp:14}.

While normal diffusion is characterized by the universal Gaussian distribution, anomalous subdiffusion is non-universal and can be due to a variety of reasons~\cite{geyer:jcp:12:MixingNormAnnomal, hoefling:rpp:anomal:13, ernst_weiss:pccp:14, meroz_sokolov:pr:15:anomal}. Detailed discussion of all possible mechanisms is out of scope of this Lecture; we shall only mention that the frequently observed mechanism, particularly in eukaryotes, is fractional Brownian motion, which corresponds to diffusion in viscoelastic-like medium~\cite{ernst_weiss:sm:12:fbm}. Macromolecular trapping with varying trapping times, which is described by the so-called continuous time random walk model, has been demonstrated to take place on short time scales~\cite{jeon_metzler:prl:11AnomalDiff}. 

\section{Effect of crowder's composition on diffusion}

\begin{figure}
	\begin{center}
	\includegraphics*[width=0.55\textwidth]{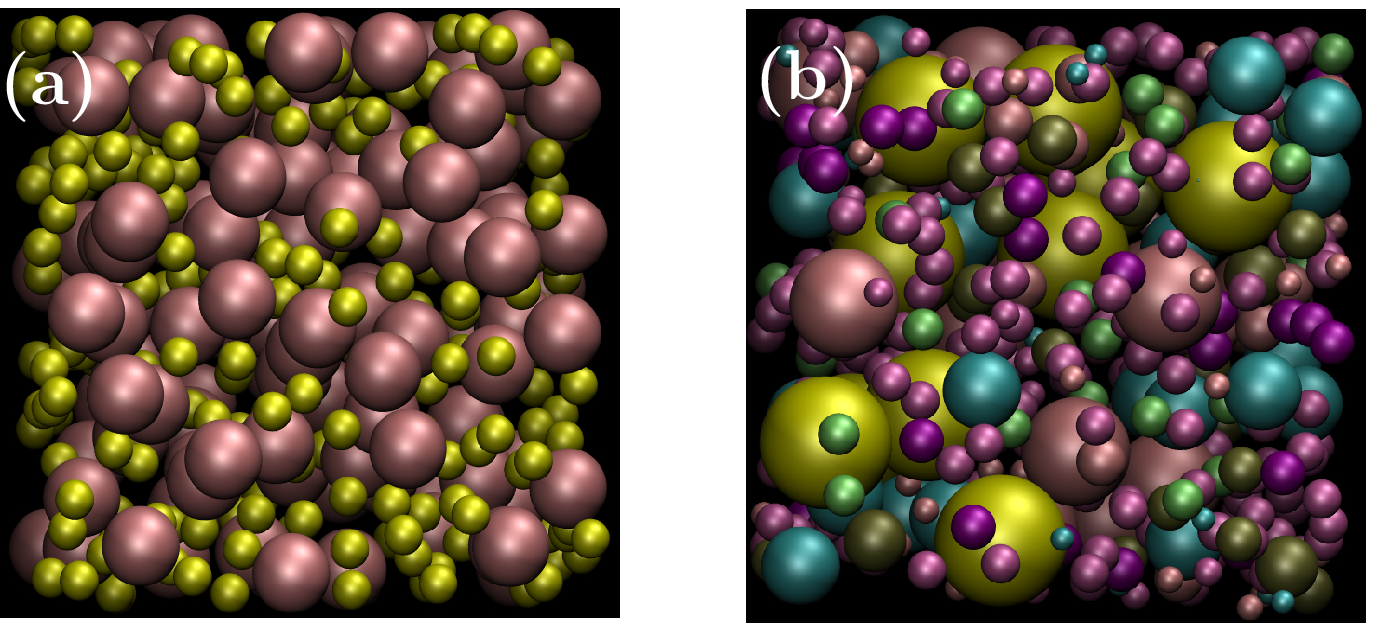}
    	\includegraphics*[width=0.38\textwidth]{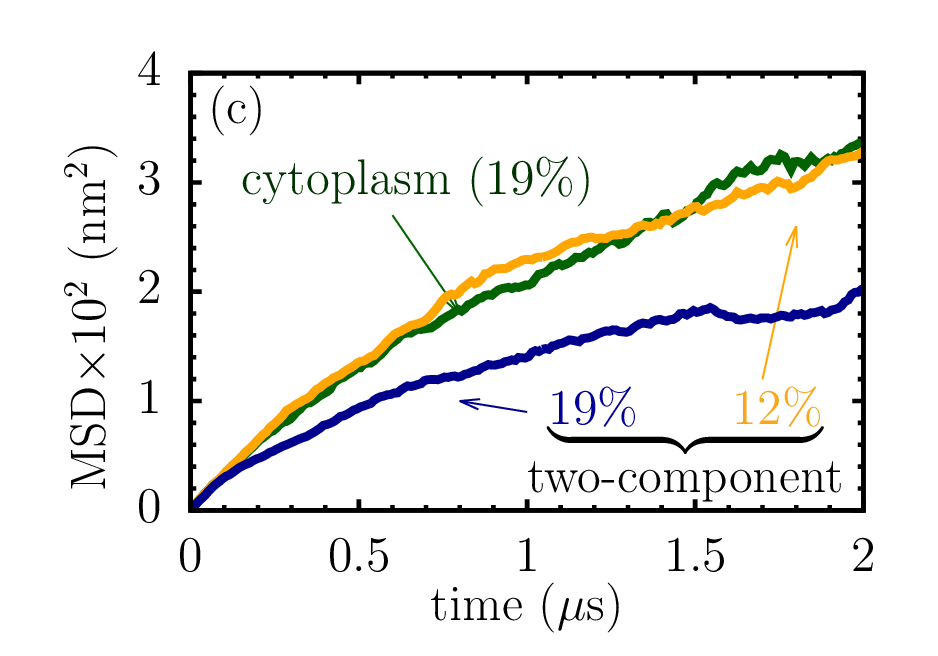}
	\end{center}
        \caption{
        \label{fig:shot}
	\figtitle{Effect of composition on diffusion.} Snapshots from Brownian dynamics simulations of (a) dense two-component system and (b) model cytoplasm. The volume fraction occupied by macromolecules is approximately $19\%$ in both cases. (c)~Comparison of the mean-square displacement (MSD) for a macromolecule of size $2.5$nm in the model cytoplasm and in a two-component system. The cytoplasm has a volume fraction of $19.2\%$, and the two-component system $18.6\%$ and $11.6\%$; the concentration of smaller molecules is $0.875$. Reproduced from \myref{kondrat:pb:15:bdcomp}}
\end{figure}

\iffindex{intracellular diffusion!effect of composition}

Living cells are characterized by constant changes of the cell constituents, causing the variation of the \emph{relative concentrations} of macromolecules. To understand how this affects diffusion, we have performed~\cite{kondrat:pb:15:bdcomp} Brownian dynamics simulations of a two-component system (\fig{fig:shot}a), where the effect of composition can be studied more systematically, and compared these results with the results for a model cytoplasm (Figure~\ref{fig:shot}b). 

\Fig{fig:shot}c demonstrates that diffusion depends sensitively on the relative concentrations of macromolecules. In particular, the (middle-time) diffusion in the cytoplasm is faster than in the two component system with the same packing fraction. Similarly, the diffusion is comparable in the cytoplasm and in the two component system (the latter with the volume fraction $12\%$), although the cytoplasm is over $50\%$ `more crowded' than the two-component system.

These results raise two interesting questions. Firstly, artificial crowders (such as dextran) are often used in \textit{in vitro} experiments (\sects{sec:diff:exp} and \ref{sec:diff:anomal}) to mimic the environment inside living cells. However, \fig{fig:shot}c suggests that such nearly monodisperse crowders may in fact be inadequate for this purpose. Secondly, the volume fraction is frequently used as a measure of crowdedness, while \fig{fig:shot}c shows that it \emph{cannot} serve as a  unique measure of how crowded a system is. Indeed, diffusion depends sensitively on molecular composition, and can in fact be faster in systems with higher volume fractions. 

\section{Conclusions and outlook}
\label{sec:concl}

We have discussed the main principles of how to model macromolecular self-diffusion, and briefly reviewed one of the most important manifestations of the crowded world inside living cells, which is a dramatic slow-down of the macromolecular diffusion as well as its anomalous behaviour. From a modelling perspective, the steric and hydrodynamic forces seem to play the key role in the reduction of the diffusion coefficient~\cite{ando:pnas:10}. However, the atomistic details of macromolecules do not seem to be of any significant importance for transport diffusion~\cite{ando:pnas:10}, which shall alleviate the computational burden of many Brownian dynamics simulations. Interestingly, the macromolecular composition affects considerably the macromolecular diffusion, which can be faster in systems with higher volume fractions~\cite{kondrat:pb:15:bdcomp}.

Many questions remain unanswered, however. I shall only briefly mention some of them.
\begin{itemize}

	\item Diverse experimental results have been reported for diffusion on long time scales. While some experiments show the slow-down of the normal diffusion~\cite{elowitz:jb:gfpmob:99, dauty_verkman:jmr:04CrowdingSmallBig, konopka:jb:gfpdiff:06}, strong evidence exists in support of the anomalous subdiffusion~\cite{weiss:bj:crowdanomal:04, banks:bj:anomal:05, ernst_weiss:sm:12:fbm, jeon_metzler:prl:11AnomalDiff}. Whether and when the anomalous diffusion turns into the normal regime is not clear. It is important to note in this context that the characteristic times of the anomalous diffusion reported in the experiments are of the order of up to a few seconds; the Brownian dynamics (BD) simulations, discussed in this Lecture, are currently unable to deal with such long time scales. On the other hand, the temporal resolution of a typical experiment is too low to access the time scales of BD simulations.

	\item We have discussed self-diffusion, which is relevant to macromolecules, since they are typically present in cells in very low `copy numbers'. However, the concentration of smaller molecules (metabolites) is often relatively high, in which case the \emph{transport} diffusion becomes a relevant process. The transport and self-diffusion can differ dramatically, but this topic has not been touched in the biophysical context.

	\item Spectacular behaviour has recently been reported by \citeauthor{parry:cell:14:glassy}~\cite{parry:cell:14:glassy}, who observed a dramatic slow-down of diffusion in cells with suppressed metabolic activity (\ie, no or few reactions taking place). The origin of this effect is not yet well understood, but it points out to an interesting inter-dependence of \textit{in vivo} reactions and diffusion. Clearly, diffusion controls the rate of (diffusion-limited) reactions, but it turns out that also reactions can influence the intracellular diffusion in a dramatic way.

	\item I hope I have convinced you that the intracellular diffusion is important and interesting; however, it is essentially reactions that make Life. Incorporating reactions in the Brownian dynamics simulations is a difficult and computationally expansive task. Although a number of multiscale and coarse-grained approaches have been introduced~\cite{wylie:jpcb:hybrid:06, kalantzis:compbiochem:09, flegg:jrsi:11, kondrat:epje:16:rds}, a well-developed reliable framework does not seem to exist. The development of such a framework, which would allow for spatially-resolved whole-cell simulations, will likely be the focus of future research activities. In combination with the advanced experimental studies, this will bring new discoveries and a better understanding of the Physics of Life.

\end{itemize}

\section*{Acknowledgment}
I am grateful to A.~Cherstvy (Uni Potsdam) and M.~D{\l}ugosz (Warsaw University) for  critical reading of this Chapter and for fruitful comments and suggestions.

\bibliography{diff}

\end{document}